\documentclass[preprint,10pt]{elsarticle}

 \usepackage{graphicx}
\usepackage{epsfig}

\usepackage{amssymb}
\journal{Physica A}

\begin{document}

\begin{frontmatter}


\title{Trapping of Continuous-Time Quantum walks on Erd\"os-R\'enyi graphs}

\author{E. Agliari}
\address{Dipartimento di Fisica, Universit\`a degli Studi di
Parma, viale Usberti 7/A, 43100 Parma, Italy}
\address{INFN, Gruppo Collegato di
Parma, viale Usberti 7/A, 43100 Parma, Italy}
\address{Theoretische Polymerphysik, Freiburg Universit\"at, Hermann-Herder-Str. 3, 79104 Freiburg, Germany}



\begin{abstract}
We consider the coherent exciton transport, modeled by continuous-time quantum walks, on Erd\"{o}s-R\'{e}ny graphs in the presence of a random distribution of traps. The role of trap concentration and of the substrate dilution is deepened showing that, at long times and for intermediate degree of dilution, the survival probability typically decays exponentially with a (average) decay rate which depends non monotonically on the graph connectivity; when the degree of dilution is either very low or very high, stationary states, not affected by traps, get more likely giving rise to a survival probability decaying to a finite value. Both these features constitute a qualitative difference with respect to the behavior found for classical walks. 
\end{abstract}

\begin{keyword}
Quantum walks \sep Trapping \sep Random graphs


\end{keyword}

\end{frontmatter}

% \linenumbers
\section{Introduction} \label{sec:intro}
%
%PAPABILI: \\I
%nt. J. Q. Comp. $0.6$\\
%Physica A $1.3$\\
%EPJ $1.6$\\
%Int. J. Mod. Phys. B $0.4$\\
%JPA $1.6$\\
%
%
Quantum walks provide a quantum extension
of the ubiquitous classical random
walks and have important applications in a broad range of
fields including solid-state physics, polymer chemistry, biology,
astronomy, mathematics and computer science \cite{kempe,supriyo,ambainis,agliari3}.  Due to their coherent nature, the behavior of quantum
walks can differ significantly from that of the classical random walks, as corroborated by measures of mixing times,
hitting times and exit probabilities of quantum walks \cite{xu1}.

The continuous-time version of quantum walks (continuous-time quantum walks, CTQWs) has been extensively studied as effective model of energy transport in molecular systems such as chromophoric light-harvesting complexes \cite{NJP,may}.
%, which represents  one of the most important classes of quantum transport \cite{NJP,may}, for example in the  \cite{engel,lee}. The role of the %environment in chromophoric systems has been widely studied \cite{grover,yang,gilmore}. 
In photosynthesis, excitation energy is absorbed by pigments present in the antennas and subsequently transferred to a reaction center where an electron-transfer event initiates the process of biochemical energy conversion. This process has been studied for decades due to its impressive efficiency (even over $90 \%$ in certain bacterial systems and higher plants), nonetheless a full description of the mechanism leading to such a remarkable efficiency has not been achieved yet \cite{bossi,caruso}. Also, the overall effect of the environment, of its (quenched) disorder and of the relative position of the reaction center are expected to play a central role \cite{NJP,sener,agliari1}.
%In particular quantum localization can emerge in the presence of disorder and is can seriously limit computational power or quantum walk %properties \cite{anderson}. Generally, the overall effect of environment and the static disorder is expected to be negative. 
%However, as we demonstrate here, in some classes of transport systems, the interaction with the environment can result in decreased quantum %efficiency which can lead to positive outcomes.
%
%
%
%
%The role of the structure has been shown to play a crucial role in the overall behavior \cite{sener}
%
Indeed, understanding how such a process works might be useful for the nano-engineering of optimized solar cells \cite{caruso}.
%Apart from the pure theoretical interest, understanding the nature of biological
%systems have direct or indirect implications to drug design, disease
%diagnosis and cure, epidemic control, and biotechnological applications.

%In particular, to analyze the survival probability in the presence of one or more absorbing site...

The success rate of an energy transfer process can be investigated by studying the interaction between a quantum walk (mimicking the rather coherent propagation of the exciton) with a reaction center (being it an impurity atom or molecule), which irreversibly traps the moving particle. Consequently, a great deal of recent theoretical work has focused on investigating essential features of basic trapping models, wherein a quantum particle moves in a medium containing different arrangements of traps. 
The trapping problem on a one-dimensional structure has already been investigated in \cite{oli_lettera,agliari2}, where distinct configurations of traps (ranging from periodical to
random) where shown to yield strongly different behaviors for the quantal mean survival probability, while classically, at long times, the exponential decay is always recovered.
In this context the case of substrates displaying random topological inhomogeneity has not yet been investigated, notwithstanding their experimental importance \cite{rand1,rand2}. Such random structures, typically modeled by Erd\"os-R\'eny (ER) graphs, have attracted a great deal of interest in the last years also due to new tools introduced for their investigation (see e.g. \cite{barra,AB}).

%In cromophoric complexes, an environment assisted quantum walk approach was also used to quantify the percentage contributions of quantum %coherence and environment-induced relaxation to the overall efficiency \cite{aspuru}

In this work we study the survival probability of a CTQW moving on an ER graph in the presence of a fixed concentration of traps randomly placed. The number of traps $M$ as well as the substrate connectivity encoded by the average number of neighbors per node $\bar{z}$, are properly tuned in order to account for their role in affecting the trapping performance. Indeed, since ER networks lack hubs and display an overall homogeneous topology, the transport properties are controlled mainly by the average degree. 
Hence, for a given realization of the system we measure the survival probability $\Pi_{M,\bar{z}}(t)$ as a function of time $t$, which, for sufficiently large systems, turns out to display a (qualitatively) robust behavior with respect to the realization of the system; however, due to the intrinsic randomness of the system (involving both the substrate topology and the trap arrangement) in order to properly outline the typical behavior, we generated several realizations (for given $M$, $\bar{z}$ and graph size $N$), over which we averaged to get $\langle \Pi_{M,\bar{z}}(t) \rangle$. Analogous calculations have been performed for the case of a classical particle modeled by a continuous-time random walk (CTRW) in order to figure out possible genuine quantum-mechanical effects. Analysis are performed both numerically and analytically relying on matrix diagonalization algorithms and on perturbation theory, respectively.

Our results highlight that at long times and when the trap concentration is small, for large size ($N \gg 1$) and intermediate degrees of dilution ($\bar{z} \sim O(N)$), both quantal and classical survival probabilities typically decay exponentially with time. However, while the classical decay rate increases monotonically with $\bar{z}$, the quantal decay rate displays a subtler dependence. Indeed, when the degree of dilution is either very large or very low, stationary states, not affected by traps, get more likely and this makes the quantal survival probability to decay to a finite value related to the number of localized eigenmodes.
As a result, given a fixed number of randomly arranged traps, in order to enhance the trapping efficiency we need to thicken the substrate connectivity, independently of the current degree of dilution, while quantum mechanically the strategy does depend on the current degree of dilution.

%Transport properties such as electrical and frictionless flow conductance on ER graphs \cite{lopez}

This paper is organized as follows. In Sec.~\ref{sec:model} we define CTRW and CTQW and we describe the substrate where they move on; then, in Sec.~\ref{sec:results} we reports our results and finally in Sec.~\ref{sec:conclusion} we discuss them and possible extensions.

\section{Coherent dynamics on random graphs} \label{sec:model}
The incoherent transport occurring over a discretizable environment can be modeled by continuous-time random walks (CTRWs) described mathematically by a master equation. However, when dealing with quantum particles at low
densities and low temperatures, decoherence can be suppressed to a
large extent. Therefore, the study of transport in this regime
requires abandoning the classical, master-equation-type formalism and adopt a quantum-mechanical oriented picture, where the local description of the complex network of molecules involved in the transport can be retained through a tight-binding approach.

Interestingly,  the CTRW picture can be mathematically reformulated to yield a quantum-mechanical Hamiltonian of tight-binding type; the procedure uses the mathematical analogies between time-evolution operators in statistical and in quantum mechanics: The result are continuous-time quantum walks (CTQWs). 

In the following we provide the formal tools for the study of both CTRWs and CTQWs.

\subsection{Graph formalism} \label{ssec:graph}
Let us consider a graph $\mathcal{G}$ made up of $N$ nodes and algebraically described by the so-called adjacency matrix $\mathbf{A}$: The non-diagonal elements $A_{ij}$ equal $1$ if nodes $i$ and $j$ are connected by a bond and $0$ otherwise; the diagonal elements $A_{ii}$ are $0$. We define the coordination number, or degree, of a node $i$ as $z_i = \sum_j A_{ij}$.

An Erd\"os-R\'enyi random graph is built as follows: Starting with $N$ disconnected nodes, every pair, say $i$ and $j$, is connected, namely $A_{ij}=1$, with probability $p$, being $0 \leq p \leq 1$; multiple connections are forbidden and the extreme cases trivially correspond to a completely disconnected graph ($p=0$) and to a fully connected graph ($p=1$).
In the limit of large size $N$, the coordination number of an arbitrary node follows a binomial distribution with average $\bar{z} = p (N-1) \approx p N$.
Due to their rigorous mathematical definition, ER graphs have been studied in details, from a topological point of view \cite{ER}, as well as for what concerns the properties of statistical mechanics models defined on them (see e.g. \cite{franz,barra}). 
In particular, the Molloy-Reed criterion for percolation \cite{a56} shows that,
in the limit $N \to \infty$ a giant component exists, namely the graph is overpercolated, if and only if $p$ is
larger than $1/N$.

Now, the Laplacian operator $\mathbf{L}$ of an arbitrary graph $\mathcal{G}$ is defined as $L_{ij} = z_i \delta_{ij} - A_{ij}$, its spectrum, i.e. the set of all $N$ eigenvalues of $\mathbf{L}$, being denoted as $0 = \lambda_1 \leq \lambda_2 ... \leq \lambda_N$; 
it follows from Ger\v{s}gorin's theorem \cite{merris1,merris2} that $\mathbf{L}$ is positive
semi-definite and, because its rows sum to $0$, $\mathbf{e_N} \mathbf{L} = 0$, where $\mathbf{e_N}$, is the row
$n$-tuple each of whose entries is $1$, therefore, the minimum eigenvalue is  $\lambda_1 = 0$ and it is afforded by $\mathbf{e_N}$.

The eigenvalues and eigenvectors of Laplacian matrix $\mathbf{L}$ (also referred to as the admittance
matrix, the stiffness matrix, or the Kirchhoff matrix) basically form the backbone
of any discussion of dynamic behavior of the networks represented by our graphs
(see for example \cite{mohar} and references therein).
In particular, the degeneracy of the null eigenvalue represents the number of disconnected components making up the whole graph (in the following we will always consider connected graphs made up of one single component), while the second smallest eigenvalue, also called spectral gap, controls the synchronization time. % \cite{donetti}.
Moreover, 
the characterizations
of eigenvectors are needed for a range of decentralized
controls and dynamical-network analysis/design applications,
including e.g., network partitioning, synchronization design, and optimal network resource allocation. Despite
such need, graph theoretic studies of the Laplacian eigenvectors
are sparse (see \cite{libro} for some reviews of these
literature), and do not provide exact general characterizations
of eigenvector-component values in terms of graph
constructs for arbitrary graphs.

In the following, we will refer to the eigenvalues and eigenvectors
of $\mathbf{L}$ as the eigenvalues and eigenvectors of $\mathcal{G}$.

To reveal the decay behavior of survival probabilities of the CTQW and CTRW we focus on systems of large size for which the spectral density of the Laplacian spectrum follows Wigner's law \cite{wigner} and converges to the semicircle distribution
\begin{equation} \label{eq:Wigner}
\rho(\lambda) = \frac{\sqrt{4 \sigma^2 - (\lambda - \bar{z})^2 }}{2 \pi \sigma^2}, \;\;\; \mathrm{if} |\lambda - \bar{z}| < 2 \sigma,
\end{equation}
where $\sigma = \sqrt{Np(1-p)}$ and $\bar{z} = pN$. Moreover, for highly diluted ($p \ll 1$) networks we can write $\sigma^2 \approx \bar{z}$ and $\rho$ can be expressed as a function of $\bar{z}$ only.

\subsection{Classical and Quantum walks on graphs} \label{ssec:walks}

Continuous-time random walks (CTRWs) \cite{weiss} are described by the following Master Equation:
\begin{equation}\label{eq:master_cl}
\frac{d}{dt} p_{k,j}(t)= \sum_{l=1}^{N} T_{kl} p_{l,j}(t),
\end{equation}
being $p_{k,j}(t)$ the conditional probability that the walker at time $t$
is on node $k$ when it started from node $j$ at time $0$. If the
walk is unbiased the transmission rates $\gamma$ are bond-independent and the
transfer matrix $\mathbf{T}$ is related to the Laplacian
operator through $\mathbf{T} = - \gamma \mathbf{L}$ (in the following we set $\gamma=1$).

We now define the quantum-mechanical analog of the CTRW, i.e. the CTQW, by
identifying the Hamiltonian of the system with the classical transfer
matrix, $\mathbf{H}=-\mathbf{T}$ \cite{farhi}.
Hence, given the orthonormal basis set $|j \rangle$, representing the
walker localized at the node $j$, we can write
\begin{equation}\label{eq:tb}
\mathbf{H} = \sum_{j=1}^{N} z_j |j\rangle \langle j| -  \sum_{j=1}^{N}
\sum_{k \sim j} |k\rangle \langle j| ,
\end{equation}
where in the second term we sum over all connected couples $k \sim j$.
The operator
$\mathbf{H}$ is just the tight-binding Hamiltonian which applies to a large class of quantum transport systems such as excitons and charges in molecular and quantum dots \cite{NJP}. 

Now, the quantum mechanical time evolution operator is defined as $\mathbf{U}(t,t_0) = \exp[-i \mathbf{H} (t-t_0)]$, so that
the transition amplitude
$\alpha_{k,j}(t)$ from state $| j \rangle$ at time $0$ to state $| k \rangle$ at time $t$ reads $\alpha_{k,j}(t) = \langle k | U(t,0) | j \rangle$, and it obeys the following Schr\"{o}dinger equation:
\begin{equation}\label{eq:schrodinger}
\frac{d}{dt} \alpha_{k,j}(t)=-i \sum_{l=1}^{N} H_{kl} \alpha_{l,j}(t),
\end{equation}
formally very similar to Eq.~\ref{eq:master_cl}. 
Then, the classical and quantum transition probabilities to go from state $| j \rangle$ to state $| k \rangle$ in a time $t$ are given by $p_{k,j}(t) = \langle k | e^{- t T} | j \rangle$ and $\pi_{k,j}(t) = |\alpha_{k,j}(t)|^2 = |\langle k | e^{- i t H} | j \rangle|^2$, respectively.

In the absence of traps and other impurities, the operators describing the dynamics of CTQWs and of CTRWs share the same set of eigenvalues and of eigenstates; denoting with $E_n$ and $|\Phi_n \rangle, n \in [1,N]$ the $n$th eigenvalue and orthonormal eigenvector of $\mathbf{L}$, we can
write
\begin{equation}
p_{k,j}(t) =  \sum_{n=1}^{N}  e^{- E_n t} \langle k |\Phi_n \rangle \langle \Phi_n | j \rangle,
\end{equation}
and
\begin{equation}
\pi_{k,j}(t) = \left| \sum_{n=1}^{N} \langle k | e^{-i E_n t} |\Phi_n \rangle \langle \Phi_n | j \rangle \right|^2.
\end{equation}

\subsection{CTRWs and CTQWs in the presence of traps}\label{ssec:Trapping}
Let us introduce a set $\mathcal{M}$ of $M$ traps placed randomly on nodes $\{ m_1, m_2, ..., m_M \}$.
In the incoherent, classical transport case trapping is incorporated into the CTRW according to
\begin{equation}
\mathbf{T} = \mathbf{T_0} - \mathbf{\Gamma} = - \mathbf{L} -\mathbf{\Gamma},
\end{equation}
where $\mathbf{T_0}$ denotes the unperturbed operator without traps
while $\mathbf{\Gamma}$ is the trapping operator defined as
\begin{equation} \label{eq:Gamma}
\mathbf{\Gamma} = \Gamma \sum_{m \in \mathcal{M}}  | m \rangle \langle m |.
\end{equation}
The capture strength $\Gamma$ determines the rate of decay for a particle located at trap site and here it is assumed to be equal for all traps. 

The transfer operator $\mathbf{T}$ is therefore self-adjoint and negative definite; we denote its eigenvalues by
$-\lambda_l$ and the corresponding eigenstates by $| \phi_l \rangle$.

The mean survival probability for the CTRW can be written as
\begin{eqnarray}    \label{eq:p}
\nonumber
\lefteqn{P_M(t) \equiv \frac{1}{N-M}  \sum_{j \notin \mathcal{M}} \sum_{k \notin \mathcal{M}} p_{kj}(t)}   \\
& & = \frac{1}{N-M} \sum_{l=1}^{N} e^{-\lambda_l t} \left| \sum_{k \notin \mathcal{M}} \langle k | \phi_l \rangle \right|^2.
\end{eqnarray}
From Eq.~\ref{eq:p} one may deduce that $P_M(t)$ attains in general rather
quickly an exponential form; furthermore, if the smallest eigenvalue
$\lambda_{\rm min}$ 
is well separated from the next closest eigenvalue, $P_M(t)$ is dominated
by 
$\lambda_{\rm min}$ 
and by the corresponding eigenstate 
$|\phi_{\rm min} \rangle$ \cite{oli_lettera,agliari2}:
\begin{equation}    \label{eq:p_asym}
P_M(t) \approx \frac{1}{N-M} e^{-\lambda_{\rm min} t} \left| \sum_{k
\notin \mathcal{M}} \langle k | \phi_{\rm min} \rangle \right|^2.
\end{equation}

As for quantum transport, in the presence of substitutional traps the system can be described by the following effective (but non-Hermitian) Hamiltonian \cite{oli_lettera}
\begin{equation} \label{eq:H}
\mathbf{H} = \mathbf{H}_0 - i \mathbf{\Gamma},
\end{equation}
where $\mathbf{H_0}$ denotes the unperturbed operator without traps. 

Due to the non-hermiticity of $\mathbf{H}$, its eigenvalues are complex and can be written as $E_l = \epsilon_l - i \gamma_l \ (l=1,...,N)$; moreover, the set of its left and right eigenvectors, $| \Phi_l \rangle$ and $\langle \tilde{\Phi}_l |$, respectively, can be chosen to be biorthonormal ($\langle \tilde{\Phi}_l |\Phi_l' \rangle = \delta_{l,l'}$) and to satisfy the completeness relation $\sum_{l=1}^N |\Phi_l \rangle \langle \tilde{\Phi}_l | = \mathbf{1}$.
Therefore, according to Eq.~\ref{eq:schrodinger}, the transition amplitude can be evaluated as
\begin{equation} \label{eq:alfa}
\alpha_{k,j}(t) = \sum_{l=1}^N e^{- (\gamma_l + i \epsilon_l)t} \langle k | \Phi_l \rangle \langle \tilde{\Phi}_l | j \rangle,
\end{equation}
from which $\pi_{k,j}(t)=|\alpha_{k,j}(t)|^2$ follows.

Of particular interest, due to its relation to experimental observables, is the mean survival probability $\Pi_M(t)$ which can be expressed as \cite{oli_lettera}
\begin{eqnarray} \label{eq:pi}
\nonumber
\lefteqn{\Pi_{M} \equiv \frac{1}{N - M} \sum_{j \notin \mathcal{M} } \sum_{k \notin \mathcal{M} } \pi_{kj}(t)} \\
\nonumber
& & = \frac{1}{N-M}   \sum_{l=1}^N  e^{-2 \gamma_l t} \left ( 1- 2 \sum_{m \in \mathcal{M}} \langle \tilde{\Phi}_l | m \rangle \langle m | \Phi_l \rangle \right )  \\
& & \mbox{} + \frac{1}{N-M} \sum_{l,l'=1}^N e^{-i(E_l - E_{l'}^{*})} \left ( \sum_{m \in \mathcal{M}} \langle \tilde{\Phi}_{l'} | m \rangle \langle m | \Phi_l \rangle \right )^2 .
\end{eqnarray}
The temporal decay of $\Pi_M(t)$ is determined by the imaginary parts of $E_l$, i.e. by the $\gamma_l$.
As shown in \cite{oli_lettera}, at intermediate and long times and for $M \ll N$, the $\Pi_M(t)$ can be approximated by a sum of exponentially decaying terms:
\begin{equation}\label{eq:pi_asym}
\Pi_{M} \approx \frac{1}{N-M} \sum_{l=1}^{N} e^{-2 \gamma_l t},
\end{equation}
and is dominated asymptotically by the smallest $\gamma_l$ values.

\section{Results} \label{sec:results}
As shown in the previous section, the long-time decay of the survival probabilities $P_M(t)$ and $\Pi_M(t)$ is controlled by the ``spectra'' $\{  \lambda_l  \}$ and $\{ \gamma_l \}$, respectively. When the capture strength is small ($\Gamma \ll 1$), some insights into such spectra can be obtained following a perturbative approach to get the correction to the unperturbed eigenvalues.
For instance, the first-order correction to $E_1$, corresponding to the (unperturbed) eigenvector $| \Phi_1^{(0)} \rangle = \mathrm{ \mathbf e_N} / \sqrt{N}$, reads as
\begin{equation} \label{eq:first}
E_{1}^{(1)} =  -i \Gamma \sum_{m \in \mathcal{M}} \left| \langle m | \Phi_1^{(0)} \rangle \right|^2 = - i \Gamma \frac{M}{N},
\end{equation}
where, recalling that the graph is connected, we applied the perturbative theory for non-degenerate eigenvalues.

\begin{figure}
 \begin{center}
\includegraphics[height=55mm]{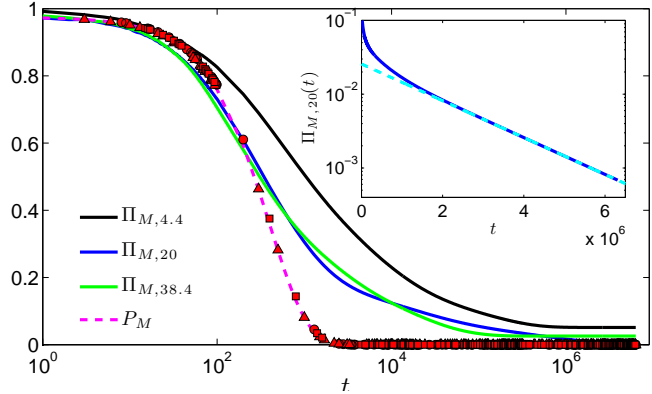}
\caption{\label{fig:special} Average Classical and Quantum survival probability as a function of time on a semilogarithmic scale-plot; while $N=40$, $M=1$ and $\Gamma=0.1$ are kept fixed, different values of $p$ are considered, that is, $p=0.11$ (darker color for the quantal probability and $\bullet$, for the classical probability), $p=0.5$ (intermediate color for the quantal probability and $\square$, for the classical probability), and $p=0.95$ (brigheter color for the quantal probability and $\triangle$, for the classical probability). The dashed, clear line represents Eq.~\ref{eq:classic} with $\Theta_c= M /N$ and is in full agreement with numerical data. In the inset we show only $\Pi_{M,20}(t)$ on a semilogarithmic scale plot in order to highlight the exponential decay; the dashed line represents the best fit with $\Theta_q \sim 10^{-5}$. 
These data refer to a particular realization of the substrate.}
\end{center}
\end{figure}

\begin{figure}
 \begin{center}
\includegraphics[height=100mm]{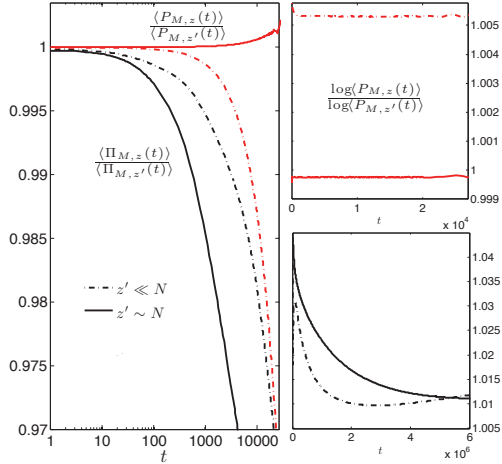}
\caption{\label{fig:compare} Ratio between the mean average classical and quantum survival probabilities $\langle P_{M,\bar{z}} \rangle$ and $\langle \Pi_{M,\bar{z}} \rangle$, pertaining to different degree of dilutions, $z$ and $z'$, for a system with $V=400$ and $M=1$. We choose as reference value $z=N/2$, while $z'$ is taken either very small or comparable with the system size, as shown in the legend. Notice that the case $p=0.5$ gives rise to relatively small survival probability in the quantal case, while classically it corresponds to an intermediate trapping efficiency in agreement with intuition.}
\end{center}
\end{figure}

Let us first focus on the classical case. From Eq.~\ref{eq:first} one can also write
\begin{equation}
\lambda_1 = \Gamma \frac{M}{N} + \mathcal{O} (\Gamma^2),
\end{equation}
and, similarly, for the remaining eigenvalues $\lambda_l = \lambda_l^{(0)} + \Gamma \lambda_l^{(1)} + \mathcal{O}(\Gamma^2)$. Now, for large system size the smallest non-null eigenvalue $\lambda_2^{(0)}$ can be estimated as $\lambda_2^{(0)} \approx \bar{z} - 2 \sigma = N p - 2 \sqrt{Np(1-p)}$ (see Eq.~\ref{eq:Wigner}), which for large average degree $\bar{z}$, is well separated by $\lambda_1$, even if corrected by a term order of $\Gamma$. Therefore, in this case the survival probability is controlled by $\lambda_{\mathrm{min}} = \lambda_1 \approx \Gamma M/N$. On the other hand, when the dilution gets lower the spectral gap also decreases and the smallest perturbed eigenvalue $\lambda_{\mathrm{min}}$ can get smaller than $\Gamma M/N$. Of course, the lower $\bar{z}$ and the smaller the expected $\lambda_{\mathrm{min}}$, in such a way that the survival probability decreases with a smaller rate. This is confirmed by numerical calculations: As shown in Fig.~\ref{fig:special} the numerical data pertaining to a given realization of the substrate are well fitted by the function
\begin{equation}\label{eq:classic}
P_{M,\bar{z}}(t) \approx \exp (- \Gamma \Theta_c t),
\end{equation}
moreover the exponent $\Theta_c$ gets lower when the link probability is drastically reduced (see Fig.~\ref{fig:compare}). 
%By realizing several substrates with fixed $M$ and $\bar{z}$ and fitting the pertaining data for the survival probability, we get the average exponent $\langle %\Theta_c \rangle$, which, as shown in Fig.~\ref{fig:espo}, increases monotonically with $p$, namely with $\bar{z}$; also notice the collapse of the rescaled %exponent $\langle \Theta_c \rangle /M$. 

%In Fig.~\ref{fig:special} we show $P_{M,\bar{z}}$, obtained from numerical simulations: different degrees of dilutions are considered, all in full agreement with %previous equations. 

The quantum-mechanical case is more subtle as in general the first eigenstate $| \Psi^{(0)} \rangle$ is not sufficient to determine the smallest perturbed eigenvalue of the complex spectrum. 

First of all, we notice that graphs displaying either high or low link dilution are more likely to present (at least one) Laplacian eigenvector displaying a large number of null entries. This is easy to see by recalling, respectively, the ``Edge Principle'' \cite{merris1,libro} and the fact that for a complete graph $K_N$ the eigenvectors show a monotonic
trend in confinement, that is, as we go toward faster modes, the eigenvector is
more localized and the region of support decreases by one node
in the graph.

%COROLLARIO che ci interessa
%Let X be an eigenvalue of G with eigenvector x such x_j = 0. 
%Let G' be the graph formed by joining some arbitrary graph H to vertex j
%of G with a single edge.
%Then \lambda is an eigenvalue of G' with eigenvector x' such that x_i' = x_i for
%i \in V(G) and x_i = 0 otherwise.

Another, intuitive, way to see this point is by noticing that the so-called Faria vectors are more likely to be Laplacian eigenvectors in the above mentioned regimes. Indeed, we recall that, in the ``valuation notation'' \cite{fiedler}, a Faria vector $| \xi \rangle$ is a vector with nonzero entries only on two vertices $i$ and $j$ with $| \xi_i \rangle = - | \xi_j \rangle =1$; moreover, a Faria vector is an eigenvector of the Laplacian of the graph $\mathcal{G}$ if and only if $i$ and $j$ are twins, i.e., if every vertex $v \notin \{ i, j\}$ is either adjacent to both $i$ and $j$ or to neither one of them. The corresponding eigenvalue is $\lambda = z_i +1 = z_j +1$ if $(i,j) \in E(\mathcal{G})$ and $\lambda = z_i = z_j$ if $(i,j) \notin E(\mathcal{G})$.
%Obviously, Faria vectors exist for arbitrarily large graphs if there is a vertex that is adjacent to at least two %vertices of degree $1$ or if 
Now, the probability that such conditions are fulfilled can be written as
\begin{eqnarray} \label{eq:faria}
\nonumber
P_F (p,N) = \sum_{k=1}^{N-2} {N-2 \choose k} (p^{2})^k [(1-p)^2]^{N-2-k}\\ 
= (1-p)^{2N-4} + [1-2(1-p)p]^{N-2},
\end{eqnarray}
which is not negligible only in the region of very high and very low dilution and for relatively small sizes; in particular, when $p$ scales like $p= \gamma/N$, being $\gamma$ a finite value, from Eq.~\ref{eq:faria} we get $P_F \sim \exp(-2 \gamma) \gamma^2/N$ for $N \gg1$. Hence, a Faria vector is more likely to be an eigenvector as $p$ approaches either $1$ or $0$, namely when
there exists relative (local) homogeneity for the two nodes corresponding to the non-null entries.

 \begin{figure}
 \begin{center}
\includegraphics[height=70mm]{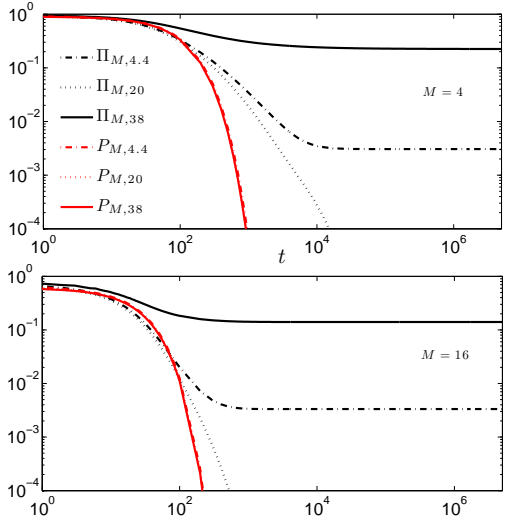}
\caption{\label{fig:V40} Comparison between the mean average classical and quantum survival probability $\langle P_{M,\bar{z}} \rangle$ and $\langle \Pi_{M,\bar{z}} \rangle$, respectively, for a system made up of $V=40$ nodes in the presence of a random distribution of $M=4$ (upper panel) and $M=16$ (lower panel) traps. Several degrees of dilutions are considered as shown in the legend, common for both panels. Notice that curves corresponding to the classical case are overlapped. Averages have been performed over $10^3$ different realizations.}
\end{center}
\end{figure}

According to formula in Eq.~\ref{eq:first}, the first-order correction is zero whenever the traps are positioned in any node other than the two twin sites; analogously, one can verify that higher order corrections are null, indeed, such highly localized states do not see the traps at all.
Consequently, the average survival probability decays to a finite value, similarly to what evidenced in \cite{agliari2} when deterministic and regular arrangements of traps were considered. The asymptotic value of $\langle \Pi_{M,\bar{z}} \rangle$ is given by the normalized number of stationary modes not intercepting the traps which, neglecting higher order terms, can be estimated by the number of disjoint Faria twins $n_F$ times the probability that they are not occupied by traps, namely $(1-M/N)^{2n_F}$ \footnote{Nonetheless, a rigorous estimate should take into account that high dilution can induce the disconnection of a subset of nodes, while at low dilution correlations among twins cannot be neglected.}. 

The above picture is confirmed by numerical results shown in Fig.~\ref{fig:special}, where the qualitative difference with respect to the classical case can also be noticed. On the other hand, for large sizes and intermediate degree of dilution, all the simulations performed recover the expected exponential decay at long times (see Eq.~\ref{eq:pi_asym}). 
%
%
%As clear from Wigner law, the spectrum distribution is most spread, i.e its support $4 \sigma$ is large, when $p = 1/2$. This is rather intuitive: when the concentration of link is very large the graph looks rather homogenoues and in the limit $p \to 1$ it recovers the fully-connected graph $K_N$. In the opposite limit of low dilution (although still over the percolation threshold) the neighborhood of nodes graph is as well rather homogeneous with a degree distribution narrowly peaked at $\bar{z}$. The regime where more variability and therefore less degeneracy is expected is just for $p=1/2$. When high degeneracy is involved for some values of $l$ first-order corrections can vanish giving rise to $\gamma_{\mathrm{min}} \sim \mathcal{O}(\Gamma^2)$ and therefore slowly decaying survival probability. This can be easily seen by looking at the case $K_N$ which can be analyzed exactly. In fact one has one non-degenerate eigenvalue $\lambda = 1$ and $N-1$-degenerate eigenvales equal to $N-1$. In this case the correction is given by...
%
%
Indeed, numerical data can be properly fitted by the function
\begin{equation}
\Pi_M(t) \approx  \exp{(- \Gamma \Theta_q t)}.
\end{equation}
%By fitting data corresponding to different realizations of the substrate we obtain an average exponent
%$\langle \Theta_q \rangle$ which, for intermediate link probability, is approximately equal to $\Gamma M/N$, consistently with previous discussions.
%More interestingly, we evidence a non monotonic behavior for the exponent as a function of $p$: a peak emerges around $p=1/2$ as expected from previous %arguments. 

%In this case we notice that $\Theta_q$ is sensitively dependent on the realization which justifies the necessity of the average.

%In quantum curves corresponding to different probabilities are resolved at intermediate times, while at short and long times they are basically indistinguishable.

Finally, we performed averages over several realizations of the underlying graph, so to get the mean average probabilities $\langle P_{M,\bar{z}} \rangle$ and $\langle \Pi_{M,\bar{z}} \rangle$.

Results are shown in Fig.~\ref{fig:V40} and Fig.~\ref{fig:compare}; in the latter figure we also plotted the ratios $\langle P_{M,\bar{z}} (t) \rangle / \langle P_{M,\bar{z}'} (t) \rangle$ and  $\langle \Pi_{M,\bar{z}} (t) \rangle / \langle \Pi_{M,\bar{z}'} (t) \rangle$. By fixing $\bar{z}=N/2$, hence corresponding to an intermediate dilution, and $\bar{z}' \ll N$ or $\bar{z}' \sim N$, we see that, classically, by reducing the dilution, the survival probability gets larger and vice versa, as expected. On the other hand, in the quantal case, both high and low dilution regimes prove to be less efficient in trapping the particles.

The slow decay we highlighted for low- and high-connectivity networks is consistent with results found previously \cite{xu} for the long time average $\bar{\chi}$ of a CTQWs embedded in ER random graphs. It is worth recalling that $\bar{\chi}$ is defined as 
\begin{equation}
\bar{\chi} = \frac{1}{N} \sum_{j=1}^N \chi_{j,j},
\end{equation}
where $\chi_{j,j}$ is the long time averaged transition probability which, classically, equals to the equal-partionend probability $1/N$, while quantum-mechanically is given by
\begin{equation}
\chi_{j,j} = \left \langle  \lim_{T \to \infty} \frac{1}{T} \int_{0}^T \pi_{j,j}(t) dt  \right \rangle.
\end{equation}
Now, $\bar{\chi}$ is found to be larger than $1/N$ and to be almost a constant value in a wide range of average degree $\bar{z}$ but increases slightly as $p$ approaches from above the percolation threshold and it increases fast when the network approaches a fully connected one where $\bar{\chi} = 1 - \mathcal{O}(N^{-1})$ \cite{xu}. 
Otherwise stated, high (and smaller) connectivities correspond to a large degree of localization hence reducing the probability to get trapped. 

\section{Conclusions} \label{sec:conclusion}
In this work we considered a continuous-time quantum walk (CTQW) propagating on Erd\"os-R\'enyi random graphs of size $N$ endowed with a tunable link probability $p$, in such a way that the average coordination number is $\bar{z}=pN$; moreover, $M$ sites extracted randomly are occupied by traps. We measured the survival probability $\Pi_{M,\bar{z}}(t)$ and we compared it to the analogous survival probability $P_{M,\bar{z}}(t)$ found for a classical continuous-time random walk (CTRW). As expected from analytical arguments, when $M \ll N$ and the graph displays a relatively large size with intermediate degree of dilution, in the long-time regime both functions typically decay exponentially with time. However, when the dilution degree is either very low or very large, while $P_{M,\bar{z}}(t)$ still decays exponentially, $\Pi_{M,\bar{z}}(t)$ can decay to a finite value due to the existence of localized eigenmodes not ``perceiving'' the traps; of course, this effect gets less likely as $M$ is increased.

We stress that the crossover evidenced in the behavior of $\Pi_{M,\bar{z}}$ for different regions of $\bar{z}$ stems from a different degree of localization exhibited by the quantum walk. Indeed, in networks with either large or very small mean degree, the quantum excitement is on average most likely to be found at the initial node \cite{xu,oli}. The reason is that the substrate topology allows the establishment of eigenstates supported by a restricted subset of nodes; this is of course a quantum-mechanical effect: the classical particle quickly reaches an equipartition condition being equally spread on the whole substrate.

As a result, while the efficiency (in terms of likelihood of trapping) of classical transport is rather robust with respect to the degree of dilution but it can be substantially reduced by cutting links, for quantum transport, improving the efficiency by adding/removing links is more a subtle operation, as it sensitively depends on the starting topology.

\bigskip
\noindent {\bf Acknowledgments} The author is grateful to A. Blumen, O. M\"{u}lken and T. Kottos for useful discussion and suggestions.
\newline
This work is supported by the FIRB grant: RBFR08EKEV.

%% References


\bibliographystyle{elsarticle-num}

\end{document}